\documentclass[pra,showpacs,preprintnumbers,amsmath,amssymb,tightenlines,twocolumn,superscriptaddress]{revtex4}

\usepackage{graphicx}
\usepackage{dcolumn}
\usepackage{bm}
\usepackage{epsfig}
\usepackage{stmaryrd}
\usepackage{color}

\newcommand{\bra}[1]{\langle\,{#1}\, |}
\newcommand{\ket}[1]{|\,{#1}\,\rangle}

\newcommand{\vek}[1]{\boldsymbol{#1}}
\newcommand{\eref}[1]{Eq.~(\ref{#1})}

\newcommand{\cref}[1]{chapter~\ref{#1}}

\newcommand{\Cref}[1]{Chapter~\ref{#1}}

\usepackage{ulem}  
\begin{document}

\title{The generalised imaging theorem: autonomous quantum to classical transitions}
\author{John S.\ Briggs}
\affiliation{Institute of Physics, University of Freiburg, Freiburg, Germany}
\email{briggs@physik.uni-freiburg.de}
\author{James M.\ Feagin}
\affiliation{Department of Physics, California State University-Fullerton, Fullerton, CA 92834, USA}
\email{jfeagin@fullerton.edu}

\begin{abstract}
The mechanism of the transition of a dynamical system from quantum to classical mechanics is of continuing interest. Practically it is of importance for the interpretation of multi-particle coincidence measurements performed at macroscopic distances from a microscopic reaction zone. 
Here we prove the generalized \emph{imaging theorem} which shows that the spatial wave function of any multi-particle quantum system, propagating over distances and times large on an atomic scale but still microscopic,  and subject to deterministic external fields and particle interactions, becomes proportional to the initial momentum wave function \emph{where the position and momentum coordinates define a classical trajectory}.
Currently, the quantum to classical transition is considered to occur via decoherence caused by stochastic interaction with an environment.
The imaging theorem arises from unitary Schr\"odinger propagation and so is valid without any environmental interaction. It implies that a simultaneous measurement of both position and momentum will define a unique classical trajectory, whereas a less complete measurement of say position alone can lead to quantum interference effects. 
\end{abstract}
\maketitle

\section{Introduction}
The development of multi-particle coincidence detectors to study fragmentation and collision processes in atomic, molecular and nuclear physics
over the past few decades has been a significant advance in experimental physics. Multi-hit detectors register the coincident positions of several particles. The position measurement is  possibly augmented by subjecting the outgoing fragments to external guiding electric and magnetic fields. This, combined with time-of-flight determination, allows measurement of position  \emph{and}  momentum of all fragments, that is, a complete characterisation of the vector correlation in position and momentum of emerging fragments. The detectors are placed at macroscopic distance from the microscopic reaction zone of atomic dimensions and so measured position and momentum must be correlated with these atomic dimensions in order that  quantum collision theory can be used to explain measured patterns in phase space. Interestingly, this is done (successfully) by using classical mechanics to trace the motion of particles from atomic to macroscopic separations. 

Such a step would appear to conflict with the description of the fragmenting complex by a many-particle quantum wavefunction which, in the vacuum of a detector, should propagate even out to macroscopic distances. That, indeed a wavefunction description is necessary, is manifest by experiments which demonstrate quantum interference even after particles have propagated to macroscopic distances.  Clearly the explanation of this dichotomy requires a consideration of the transition from quantum to classical motion. Equally clearly one can expect the answer to lie in a semi-classical treatment. However, although the construction of the semi-classical wavefunction involves  classical trajectory information and hence can
 explain interference, e.g.\ as due to contributions from two alternative trajectories, it does not directly explain how purely classical mechanics can be used to track the motion of
 particles from reaction zone to detector. That is, \emph{how does this classical behaviour arise when the particles are still governed by a wavefunction?}

Here we demonstrate that features of the transition to classical motion appear naturally from a purely quantum propagation, including propagation in the presence of mutual particle interactions and applied external fields, typically over times and distances which are microscopic.  In particular, 
we show that the locus of points of equal quantum probability defines a classical trajectory. Hence classical motion of detected particles will be inferred at all distances at which detectors are located in practical experiments.

We emphasise that these classical aspects arise solely from the unitary Schr\"odinger propagation of the system wave function without coupling to a quantum environment.  After propagation to distances which are still on the nanoscale, it is shown that  classical motion is  encoded in the wave function itself in that  \emph{each and every point} of the wave function at different times is connected by a classical relation between coordinates and momenta. This is a far more powerful statement than Ehrenfest's theorem involving  only averages.

The essential ingredient of our proof is simply to unify two well-known aspects of quantum dynamics whose connection hitherto has not been recognised. The first is that classical motion arises simply from the result that, for large asymptotic times and distances, the quantum propagator can be approximated by its semiclassical form. This form is decided purely by classical mechanics through the classical action function. The semiclassical approximation is  well developed and basic to the whole field of semiclassical dynamics and chaos in quantum theory \cite{semicl, Brumer}.

The second aspect is the imaging theorem (IT) known from scattering theory.
The IT has a long history, beginning with the work of Kemble in 1937  \cite{Kemble} who wished to identify the momentum of a freely-moving  collision fragment arriving at a detector. It has been derived for free motion several times since then, in connection with multi-particle quantum collision theory \cite{ITfree, BriggsFeagin_IT}.  Recently we have extended its validity to the extraction of collision fragments by constant electric and magnetic fields \cite{BriggsFeagin_IT}. The IT relates the coordinate space wave function at large distances from a collision region to the momentum space wave function at the boundary of the collision region and involves a  classical relation between position and momentum variables. Again, however, the precise connection to the semiclassical propagator has not been recognized.

Here we combine these two aspects in that we use the results of semiclassical quantum theory to generalise the IT to (non-relativistic) motion under the influence of arbitrary external laboratory fields and particle interactions. Our generalised IT shows that the spatial wave function of any quantum system 
of particles propagating over macroscopic distances and times becomes proportional to the initial momentum wave function, where the position and momentum coordinates are related by classical mechanics. Most importantly, this implies that the probability to measure a particle at a given position at a certain time is identically equal to the probability that it started with a given momentum at an earlier time and has moved according to a classical trajectory. 
If an experiment is designed to define a final position \emph{and} momentum 
the trajectory is unique. If only position is measured 
 then more than one trajectory  can contribute to the wave function and give rise to interference, 
as is well documented with neutron and atom interferometry \cite{RauchWerner, Pritchard_RMP, Greenberger}. 
 Hence, a detection, whether showing interference or not, will infer classical behaviour of the 
quantum system \emph{without any environmental interaction whatsoever.} In short, an observer would conclude that the motion is classical despite it being governed by the Schr\"odinger equation. 

Of course the quantum to classical transition is often connected with more philosophical arguments as to the meaning of the wave function, whether a particle is a wave, does the wave function ``collapse" during measurement, etc. Here we adopt a straightforward interpretation of measurement. A particle is always a particle. The wave function is simply an ``information field" whose squared modulus of the amplitude gives, via Born's rule, the probability to detect a particle at a certain position or momentum.  
Recently  \cite{BriggsFeaginGer}, we have emphasised that, defining the probability of detection according to this simple interpretation of the measurement process, gives all standard results of many-body scattering theory used successfully to reproduce the results of countless experiments on fragmentation processes in atomic, molecular and nuclear physics. From the IT, the results of detection of different particles at different phase space points will be compatible
with their classical motion, even though describable by a quantum wave function.

The plan of the paper is as follows. In section II we derive the generalised IT. 
in section IIIA, we give an estimate of the distances and times from the interaction region at which the IT becomes valid. Then, in section IIIB, we present the interpretation of interference experiments in terms of the IT wavefunction. Finally, in section IIIC, we discuss implications of the results of this paper for the widely-accepted explanation of the quantum to classical transition as due to the decoherence phenomenon.

\section{ The semi-classical propagator and the imaging theorem}
To be precise, we consider a system of $n$ quantum particles described by $3n$ dimensional position
vector $\vek r$ or momentum vector $\vek p$. 
Quantum particles are to be understood as material particles whose size and energy are sufficiently small that their motion must be described by quantum mechanics.
The particles interact in a volume of microscopic dimensions and emanate, at time $t = t_i$, from this volume of interaction with a momentum distribution described by the wave function $\tilde\Psi(\vek p, t_i)$. There follows propagation, usually under the influence of external forces and possible lensing systems and long range mutual interactions, to a time $t = t_f$ and a point of detection  $\vek r(t_f) = \vek r_f$ at macroscopic distances from the reaction volume. 
The corresponding state of the system $\ket{\Psi(t_i)}$ propagates in time according to
$\ket{\Psi(t)} = U(t,t_i) \,\ket{\Psi(t_i)}$,
where $U(t,t_i)$ is the time-development operator. 
Projecting onto an eigenstate $\bra{\vek r_f}$ of final position $\vek r_f$ and inserting a complete set of momentum eigenstates, this propagation is expressed in terms of wave functions as
\begin{equation}
\Psi(\vek r_f,t_f)  = \int  d\vek p \, \tilde K(\vek r_f,t_f; \vek p,t_i) \,  \tilde\Psi(\vek p, t_i),
\label{Psi_mixed}
\end{equation}
where $\tilde K(\vek r_f,t; \vek p,t_i) = \bra{\vek r_f}U(t,t_i)\ket{\vek p}$ is the mixed coordinate-momentum propagator. 
 
\begin{figure}[b]
\includegraphics[scale=.20]{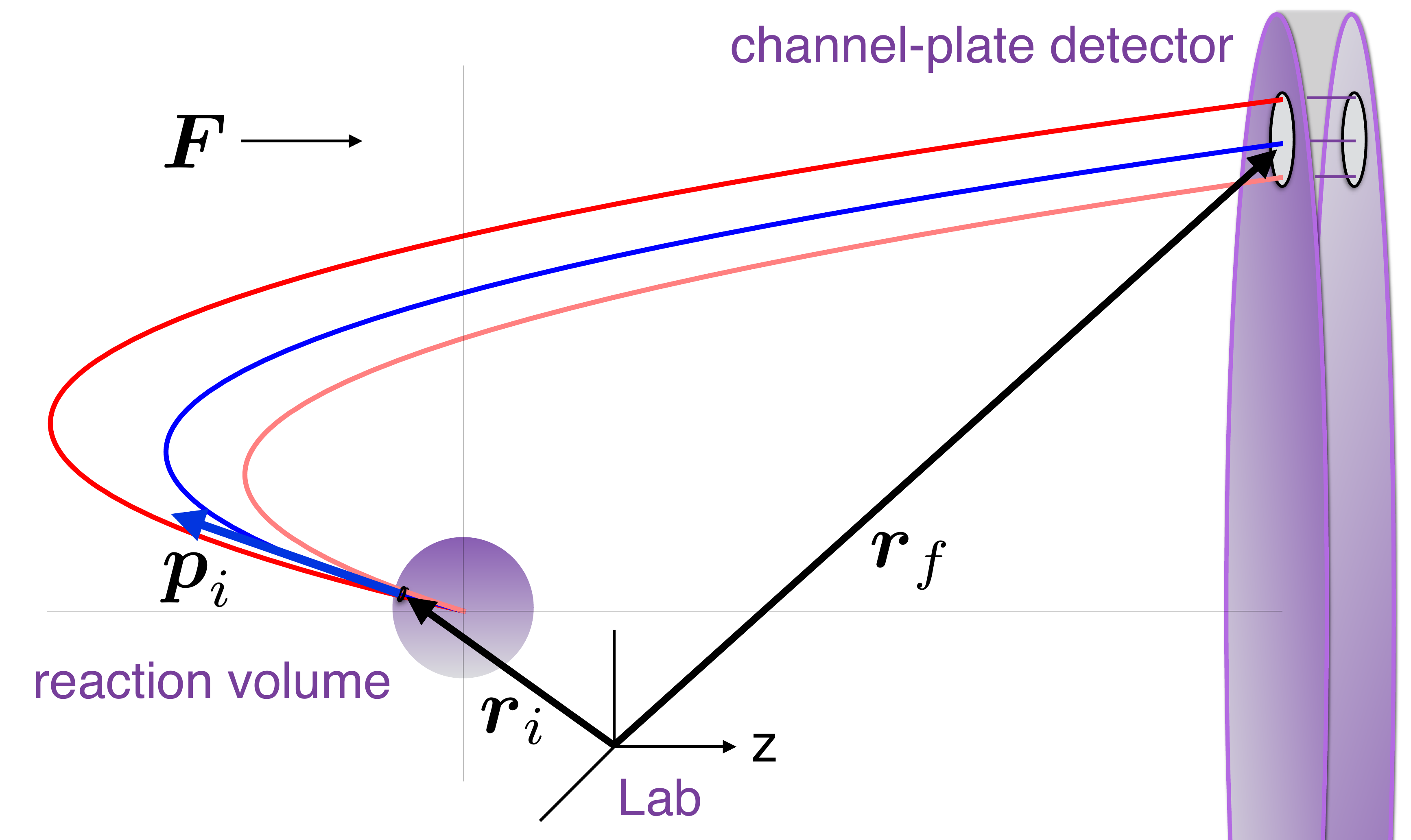}
\caption{\label{fig0} Extraction and $4\pi$ detection of a single reaction fragment by a uniform force $\vek F$ onto a channel-plate detector. The three curves represent a bundle of classical trajectories starting at $\vek r_i$ with initial momentum near $\vek p_i$ and ending near the point $\vek r_f$ on the face of the distant detector at various times $t_f$. }  
\end{figure}

In principle, the propagator is described exactly by a Feynman path integral involving the action $ \tilde S(\vek r,t; \vek p,t_i)$.
Instead, let us assume that propagation has proceeded to phase space points 
$\vek r, \vek p,t$ where the action $\tilde S$ is large
compared to $\hslash$. 
For larger times we may approximate the propagator by the corresponding  semiclassical propagator in which the \emph{classical} action $ \tilde S_c(\vek r,t; \vek p,t_i)$ appears. The boundary of this transition zone from quantum to classical action is designated by  classically conjugate variables $\vek r_i, \vek p_i,t_i$. 
This semiclassical propagation is depicted in Fig.\ \ref{fig0}.

The semiclassical mixed propagator is given by \cite{Brumer}
\begin{eqnarray}
\label{SemiK}
 \tilde K_{sc}(\vek r_f,t_f; \vek p,t_i) &=& (2\pi i \hslash)^{-3n/2} \, 
 	\left|\det \frac{ \partial^2 \tilde S_c}{\partial \vek r_f \partial \vek p} \right|^{1/2}  \nonumber \\
	&\times& \exp\left(\frac{i}{\hslash} \tilde S_c(\vek r_f,t_f; \vek p,t_i)  \right).
\end{eqnarray}
In the following, for simplicity, we will consider a single trajectory in the $3n$-dimensional space where individual particle trajectories are fully defined by measurement of appropriate positions and momenta. For this reason we suppress a possible set of Maslov phases in the equation above. The question of measurement of interference among different trajectories is discussed in section III.

We relate the mixed action $\tilde S_c(\vek r_f,t_f; \vek p,t_i)$ to the action in coordinate space
$ S_c(\vek r_f, t_f; \vek r ,t_i)$ by the  Legendre transformation
\begin{equation}
\tilde S_c(\vek r_f,t_f; \vek p,t_i) = S_c(\vek r_f, t_f;\vek r, t_i)  + \vek  p \cdot (\vek r - \vek r_i),
\end{equation}
where here $\vek r$ is considered a function of $\vek r_f$ and $\vek p$ and the times $t_f, t_i$. When the propagator \eref{SemiK} is substituted in Eq.\ (\ref{Psi_mixed}),
 the stationary phase of the integral is defined by 
 $\partial \tilde S_c/\partial \vek p = \vek r  - \vek r_i \equiv 0$,
and the root of this equation defines a stationary phase point $\vek p \to \vek p_i$.

Evaluating the integral in the stationary-phase approximation \cite{semicl} gives
\begin{eqnarray}
\label{PsiSPA}
\Psi(\vek r_f,t_f)  &\approx& (2\pi \hslash)^{3n/2} \, \left|\det \frac{\partial^2  \tilde S_c}{\partial \vek p_i \partial \vek p_i} \right|^{-1/2} \nonumber \\ 
&\times& \tilde K_{sc}(\vek r_f,t_f; \vek p_i,t_i) \,  \tilde\Psi(\vek p_i, t_i).
\end{eqnarray}
Here the determinant of the Hessian $\partial_{\vek p_i, \vek p_i }^2  \tilde S_c$ combines with the determinant in Eq.\ (\ref{SemiK}) to give the familiar
Van Vleck determinant of the Jacobian $\partial_{\vek r_f} \vek p_i= -\partial^2_{\vek r_f, \vek r_i} S_c$ according to
\begin{equation}
\left|
\det \frac{\partial^2  \tilde S_c}{\partial \vek p_i \partial \vek p_i} \right|^{-1/2} \left|\det \frac{ \partial^2 \tilde S_c}{\partial \vek r_f \partial \vek p_i} \right|^{1/2} 
= \left|\det \frac{\partial^2 S_c}{\partial \vek r_f \partial \vek r_i} \right|^{1/2}. 
\label{VVdet}
\end{equation}
Thus we obtain for a single trajectory of an $n$-particle system the asymptotic wave function
 \begin{equation}
 \label{Psixftf}
 \Psi(\vek r_f,t_f)  \approx (2\pi \hslash)^{3n/2}
 	K_{sc}(\vek r_f,t_f; \vek r_i,t_i) \,  \tilde\Psi(\vek p_i, t_i),
\end{equation}
where the coordinates $\vek r_f$ and $\vek r_i$ are connected by the classical trajectory defined by $ \vek p_i$ and
\begin{eqnarray}
\label{SemiKc}
K_{sc}(\vek r_f,t_f; \vek r_i,t_i) &=& (2\pi i \hslash)^{-3n/2} \, 
 	\left|\det \frac{ \partial^2 S_c}{\partial \vek r_f \partial \vek r_i} \right|^{1/2}  \nonumber \\
	&\times& \exp\left(\frac{i}{\hslash} S_c(\vek r_f,t_f; \vek r_i,t_i)  \right),
\end{eqnarray}
is the semiclassical coordinate propagator, form identical to the mixed propagator in Eq.\ (\ref{SemiK}) with an amplitude given by the Van Vleck determinant in Eq.\ (\ref{VVdet}).

The Van Vleck determinant also defines the classical trajectory density $d \vek p_i/d \vek r_f$ of finding the system in the volume element $d\vek r_f$ given that it started with a momentum $\vek p_i $ in the volume element $d \vek p_i$ (see Gutzwiller \cite{semicl}, chap.\ 1). 
Thus, inserting Eq.\ (\ref{SemiKc}) in Eq.\ (\ref{Psixftf}) and taking the modulus squared, one obtains
\begin{equation}
\label{VVtheorem2}
|\Psi(\vek r_f, t_f)|^2 \approx \frac{d \vek p_i}{d \vek r_f} \, |\tilde\Psi(\vek p_i, t_i)|^2,
\end{equation}
which has a wholly classical interpretation.
 A set of classical particles have momenta distributed with probability density $ |\tilde\Psi(\vek p_i, t_i)|^2$ and move along classical trajectories. This probability is multiplied by
the classical trajectory density  $d \vek p_i/d \vek r_f$.  The product gives then the classical probability density of arriving at $\vek r_f$ represented by $|\Psi(\vek r_f,t_f)|^2$.  Quantum mechanics merely furnishes the initial momentum distribution. 
This equation is trivially re-written to equate probabilities, i.e.
\begin{eqnarray}
 |\Psi(\vek r_f,t_f)|^2 \,d \vek r_f  \, \approx \,   |\tilde\Psi(\vek p_i, t_i)|^2 \, d \vek p_i.
\label{ITprob}
\end{eqnarray}
which shows that the locus of points of equal detection probability is exactly the classical trajectory.
One can also view this result as the quantum generalisation of the classical trajectory density.
Namely,
\begin{eqnarray}
\frac{d \vek p_i}{d \vek r_f} =  \left|\det \frac{\partial^2  S_c}{\partial \vek r_f \partial \vek r_i} \right|  \, \approx \,  \frac{|\Psi(\vek r_f,t_f)|^2}{|\tilde\Psi(\vek p_i, t_i)|^2}.
\label{VVtheorem}
\end{eqnarray}
Eqs.\ (\ref{Psixftf}), (\ref{VVtheorem2}), and (\ref{VVtheorem}) embody the generalised IT and are the main results of this paper. 
They justify using classical trajectories to interpret measurements on quantum particles as discussed in more detail below. 
Although direct momentum measurement is less common, Eqs.\ (\ref{Psixftf}), (\ref{VVtheorem2}) and (\ref{VVtheorem}) are readily inverted to describe a detection of the system with momentum $ \vek p_f$ given that it started near $\vek r_i $, e.g.\
\begin{equation}
\label{VVtheorem3}
|\tilde \Psi(\vek p_f, t_f)|^2 \approx \frac{d \vek r_i}{d \vek p_f} \, |\Psi(\vek r_i, t_i)|^2.
\end{equation}

\section{Discussion}
\subsection{Where is the IT valid?}
The extent of the emergence of classical motion described by the IT is seen by direct comparison of \eref{Psixftf} and the exact \eref{Psi_mixed}.
In the latter the relation between the non-commuting variables $\vek r$ and $\vek p$ is nondeterministic in that the spatial wave function at position $\vek r_f$and time $t_f$ is given by a transform at time $t_i$ of the momentum wave function involving integration over all possible values of  $\vek p$. By contrast, 
the IT of \eref{Psixftf} expresses the result that the asymptotic wave function at $\vek r_f$  and $t_f$ is given simply by the semiclassical wavefunction for the system emerging at time $t_i$ from the point $\vek r_i$ but weighted by the \emph{exact} momentum wave function at time $t_i$ of particles with momentum $\vek p_i$, where $\vek r_i, \vek p_i$ and $\vek r_f$ are classical variables connected deterministically by the classical trajectory.
This connection can, in principle, be continued all the way in to the edge of the transition zone.
Hence it remains to consider how large is the limit of the zone beyond which classical motion is manifest and the IT is valid.

Let us consider the absolutely simplest case, that of a single particle of mass $m$ undergoing free motion in one dimension 
described initially by a Gaussian of width $\sigma$ given by
\begin{eqnarray}
\label{2wfns}
\Psi(z,t_i) &=& (\pi\sigma^2)^{-1/4} e^{-z^2/(2\sigma^2)}, \nonumber \\
\tilde\Psi(p) &=&  \left(\frac{\sigma^2}{\pi \hslash^2}\right)^{1/4} e^{-p^2 \sigma^2/(2\hslash^2)}.
\end{eqnarray}
The classical action is $S_0 = m (z_f-z_i)^2/(2t)$ with $t \equiv t_f - t_i$.  
The initial momentum is given by 
$p_i = -\partial S_0/\partial z_i = m(z_f - z_i)/t$ so that $z_f = z_i + p_i \, t/m$, as desired.
The Van Vleck determinant of \eref{SemiKc} is $dp_i/dz_f = m/t$.
In this case, the semiclassical propagator is also the exact quantum propagator.
Since all $z_i$ are of microscopic size and the $z_f$ are considered macroscopic, it suffices if one takes, as assumed in experiment, $z_i = 0$. 
Then the IT \eref{Psixftf} takes the standard form \cite{ITfree,BriggsFeagin_IT},
\begin{equation}
\label{onedIT}
\Psi(z_f, t_f) = \left(\frac{m}{i t}\right)^{1/2} \exp\left[i \frac{m z_f^2}{2\hslash t}\right]  \tilde \Psi(p_i).
\end{equation} 

For $t_f > t_i$ the initial spatial wave function propagates freely in time and has the exact form
\begin{eqnarray}
\label{PsiFree}
\Psi(z_f, t_f) &=& \left(\frac{\sigma^2}{\pi}\right)^{1/4} \left(\sigma^2 + \frac{i\hslash t}{m}\right)^{-1/2} \nonumber \\
&\times& \exp{\left[-\frac{z_f^2}{2} \frac{\sigma^2- i\hslash t/m}{\sigma^4+\hslash^2 t^2/m^2}\right]}.
\end{eqnarray}  
The IT condition emerges in the limit of large times which here corresponds to $\hslash t/m \gg \sigma^2$. Then, since $p_i \equiv m z_f/t$ from the classical condition, the spatial wave function evolves into
\begin{equation}
\label{PsiFreeIT}
\begin{split}
\Psi(z_f, t_f) &\approx \left(\frac{m}{ i t}\right)^{1/2} \exp\left[i \frac{m z_f^2}{2\hslash t}\right] \\ 
&\times \left(\frac{\sigma^2}{\pi \hslash^2}\right)^{1/4} \exp\left[ - \frac{(m z_f/t)^2 \sigma^2}{2 \hslash^2}\right]\\
&= \left(\frac{m}{i t}\right)^{1/2} \exp\left[i \frac{m z_f^2}{2\hslash t}\right] \tilde \Psi(p_i),
 \end{split}
\end{equation}
that is, exactly the IT of \eref{onedIT}.

\begin{figure}[b]
\includegraphics[scale=.25]{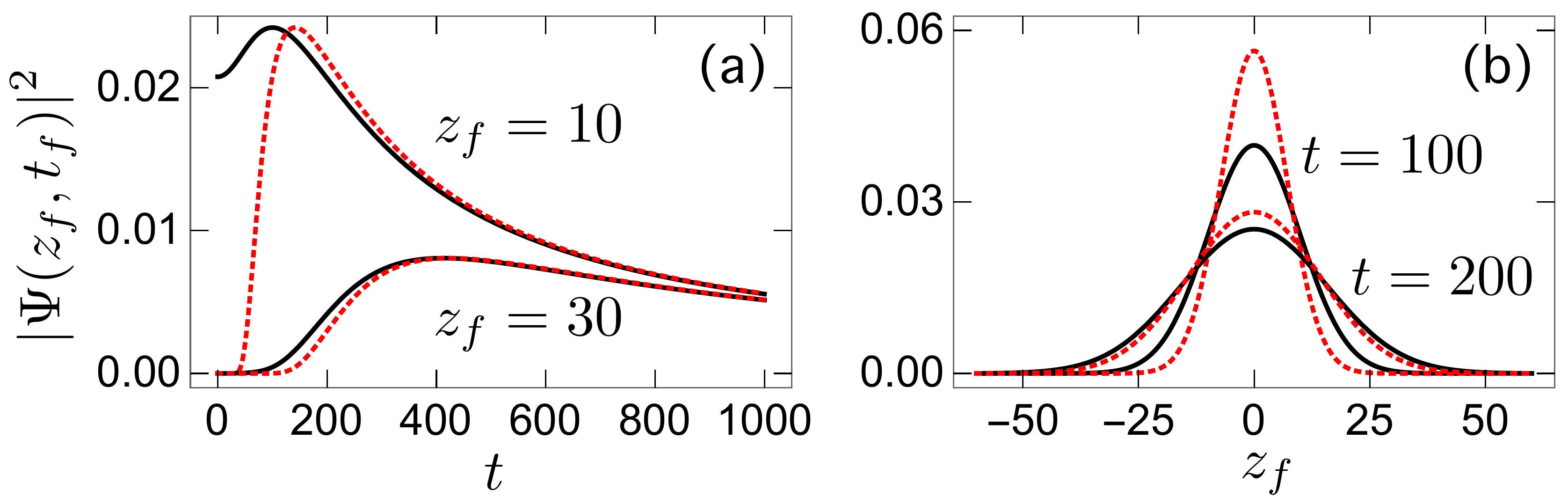}
\caption{\label{fig1} Convergence to the IT limit demonstrated by the free motion of a $1D$ Gaussian. Panel (a): the probability density as a function of time for two \emph{fixed} detector positions $z_f$.  Panel (b): the probability density as a function of $z_f$ for two different times $t$. The solid black and dotted red curves show the exact density and the IT result, respectively.}
\end{figure}

We demonstrate convergence of the exact wavefunction \eref{PsiFree} to the IT result \eref{PsiFreeIT} in Fig.\ \ref{fig1}. We take the case where $\hslash = m = 1$, which in atomic units (a.u.) corresponds to an electron wavepacket. The width $\sigma$ is taken to be $10$ a.u., or roughly
$5 \times 10^{-10} \, \mbox{m}$. Panel (a) shows the probability density $|\Psi(z_f,t_f)|^2$, 
 exact from Eq.\ (\ref{PsiFree}) and the IT approximation from Eq.\ (\ref{PsiFreeIT}), as a function of time for two fixed detector positions of $z_f = 10$ and $30$ a.u.\ which, although microscopically small, already illustrate convergence since one sees that for $z_f$ of only $30$ a.u.\ the IT result agrees closely with the exact result.  In both cases the  $m/t$ dependence of the classical density is clearly evident. In panel (b) of Fig.\ \ref{fig1}  the convergence is further demonstrated by the alternative plots of the probability density as a function of $z_f$ for two different times $t = 100$ and $200$ a.u.. 
Besides the familiar spreading of the wavefunction with time, one sees that for $z_f > 10$ and $t > 200$ there is convergence to the IT result. 

It is instructive to generalize this example to include accelerated motion due to a constant force $F$ acting along the positive $z$ axis, an example  relevant to electric-field extraction ($F = q E$)  and detection \cite{BriggsFeagin_ITII} as well as atom interferometry in a gravitational field ($F = mg$) \cite{Greenberger}.
The classical action is given by \cite{semicl}
\begin{equation}
\label{K_F}
S_F(z_f,t; z_i, t_i) = F t \, z_f - \frac{F^2 t^3}{6m} + \frac{m}{2t} \left[z_f - z_i -  \frac{F t^2}{2m} \right]^2,
\end{equation}
where again $t \equiv t_f - t_i$. 
Now the initial momentum is given by 
$p_i = -\partial S_F/\partial z_i = m[z_f - z_i -  F t^2/(2m)]/t$ so that $z_f = z_i + p_i \, t/m + F t^2/2m$, as desired.
Nevertheless, the Van Vleck determinant of \eref{SemiKc} is unchanged, $dp_i/dz_f = m/t$.
The semiclassical propagator \eref{SemiKc} is again also the exact quantum propagator and reduces to the free-particle propagator for $F = 0$.

This constant-force action $S_F$ is essentially a coordinate-translated version of the free-particle action $S_0$. Hence, the accelerated state evolves as a Galilean-like boost of the free propagation description and takes on the exact form \cite{BriggsFeagin_ITII}
\begin{equation}
\label{PsiF}
\Psi_F(z_f,t_f) =  e^{i F t \, z_f/\hslash - i F^2 t^3/(6m \hslash)} \, \Psi(z_f-Ft^2/(2m),t_f).
\end{equation}  
The limit $\hslash t/m \gg \sigma^2$ is therefore obtained from \eref{PsiFreeIT} with the substitution $z_f \to z_f-Ft^2/(2m)$. The IT is generalised with minor rearrangement accordingly,
\begin{equation}
\label{PsiFIT}
\begin{split}
\Psi_F(z_f,t_f) & \approx e^{i F t \, z_f/(2\hslash) - i F^2 t^3/(24m \hslash)}  \\
&\times  \left(\frac{m}{i t}\right)^{1/2} \exp\left[i \frac{m z_f^2}{2\hslash t}\right] \tilde \Psi(p_i),
 \end{split}
\end{equation}  
where now $p_i = m[z_f -  F t^2/(2m)]/t$. This result differs from the force-free IT  \eref{PsiFreeIT} by only an $F$-dependent phase in agreement with \eref{VVtheorem2}, since the Van Vleck determinant is the same factor $m/t$ in both cases.

\begin{figure}[b]
\includegraphics[scale=.25]{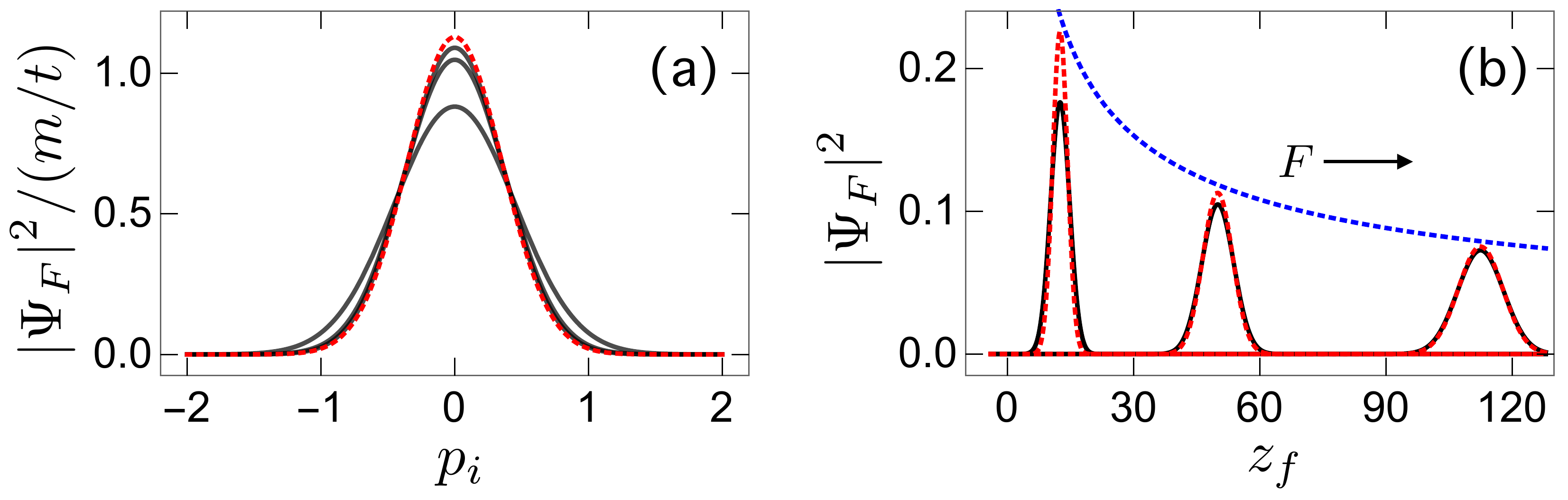}
\caption{\label{fig2} Convergence to the IT limit demonstrated by the accelerated motion of a $1D$ Gaussian. Panel (a): the probability density divided by the classical density $m/t$ as a function of $p_i$ for three different times $t$ increasing bottom to top. Panel (b): the probability density as a function of $z_f$ for the same three different times increasing left to right. The solid black and dotted red curves show the exact density and the IT result, respectively.
 The dotted blue asymptote in panel (b) shows $m/t$ as a function of $z_f$. }  
\end{figure}

We demonstrate convergence of the accelerated-state exact wavefunction \eref{PsiF} to the IT result \eref{PsiFIT} in Fig.\ \ref{fig2}. Again we consider the case of an electron wavepacket and variables are plotted in a.u.\ with $\hslash = m = 1$. Here, however, in order to emphasize better the acceleration, we take $\sigma = 2$ a.u.\ and assume an extraction field of the (unrealistically large) amplitude $F =1$ a.u.\ corresponding to $\sim \! 5 \times 10^{11} \, \mbox{volt/m}$.  

In panel (a) of Fig.\ \ref{fig2} we show the probability density $|\Psi_F(z_f,t_f)|^2$ divided by $dp_i/dz_f = m/t$. This is plotted as a function of $p_i = m[z_f -  F t^2/(2m)]/t$ for three different times $t = 5, 10,$ and $15$ a.u.\ from bottom to top, which from \eref{PsiFIT} should converge to the initial momentum probability density $|\tilde \Psi(p_i)|^2$ as $t \to \infty$. That this convergence is indeed rapid is shown by the near agreement of the exact curve for $t = 15$ with this asymptotic limit. 
In panel (b) we show the corresponding propagation in time of the probability density as a function of $z_f$ for the same three times. The rapid convergence to the IT result is again clearly evident. As $t$ increases, the wavefunction spreads and therefore drops in height, as required to conserve probability. However, one sees that the drop in height follows asymptotically the classical density as a function of $z_f$, $m/t = \sqrt{m F/(2z_f)}$, which emphasises the classical interpretation of quantum probability conservation.

The condition for validity of the IT is that the length $(\hslash t/m)^{1/2}$ be greater than the length $\sigma$. Then let us define the beginning of the transition zone to be at the position $z_i = (\hslash t_i/m)^{1/2} = f\,\sigma$, where $f$ is a number much larger than unity. Taking the mean momentum of the wave packet components to be given by 
$\bar p = \hslash/\sigma$ gives a mean kinetic energy of $\bar E = \hslash^2/(2m\sigma^2)$. The condition that the semiclassical propagator is valid is that $\bar E\,t_i \gg \hslash$. Substituting for $t_i$ gives the condition $f \gg \sqrt 2$, which is essentially the same as the IT validity condition. Note that for fixed $\sigma$ the joint condition is independent of the mass of the particle.

As realistic examples consider the dissociation of the $H_2$ molecule into two $H$ atoms or the ionisation of an electron from the ground state of the hydrogen atom. In both cases  the spatial wave packet produced will have an initial width of $\sigma \sim 1$ Bohr radius, or 1 atomic unit (a.u.) of length. If one takes the large value of $f = 100$, then the transition zone for validity of the IT begins already at the microscopic distance $z_i \sim 100$ a.u. The corresponding time for the electron to reach $z_i$ is $t_i = 10^4$ a.u.\ and for the proton, with a mass $\sim \!  10^3$ larger is 
$\sim  \!  10^7$ a.u. However, since $1$ a.u.\ of time is $\sim \!  10^{-17}\, \mbox{s}$, these are microscopic times. 
Careful time of flight experiments may be able to map this quantum to classical transition \cite{Helm2014}.

The satisfaction of the limit for the validity of the IT already for microscopic times and distances implies that the quantum wave function assumes a form leading to a classical interpretation of the results of measurement of position and momentum and correspondingly for any observable quantity composed of them. As depicted in Figs.\ \ref{fig1} and \ref{fig2}, the spatial wave function spreads of course as a function of $z_f$ as $t$ increases. However, considered as a function of $p_i$, it remains of microscopic extent. This is the effective localisation along classical trajectories occurring as a consequence of unitary Schr\"odinger propagation beyond the transition zone. As a corollary of \eref{ITprob} one has also the result, true generally, that 
\begin{equation}
|\Psi(\vek r'_f,t'_f)|^2 \, d\vek r'_f = |\Psi(\vek r_f,t_f)|^2 \, d\vek r_f
\end{equation}
for any two points along the classical trajectory, where here the volume elements are defined by the density of classical trajectories and the Van Vleck determinant according to $d\vek r_f = d\vek p_i/|\det \partial^2 S_c/\partial \vek r_f \partial \vek r_i|$. 
This relation is evident in Fig.\ \ref{fig2} as the large-$t$ limit is approached.

\subsection{Multi-path Interference}
As described in the Introduction, with multi-hit coincidence detectors it is common to employ constant electric and magnetic fields to guide charged particles from a microscopic reaction zone to the detector plates at macroscopic distances away. By extrapolating measured position and momenta back to the reaction zone \emph{using classical mechanics}, comparison is made to standard multi-particle scattering theory, derived on the assumption that asymptotic motion is free. The justification for this is given by Eq.\ (\ref{VVtheorem2}) which shows that the probability of position measurement at $\vek r_f$ is given by a wholly classical connection to an initial momentum $\vek p_i$. Indeed, as shown in \cite{BriggsFeagin_IT} the T-matrix element in momentum representation is proportional to the momentum wavefunction $\tilde \Psi(\vek p_i)$ at the exit from the reaction zone. The subsequent propagation to the detector is decided by the classical density of trajectories factor $d \vek p_i/d \vek r_f$.

Quite what is observed depends upon the measurement made. If the full vector position and momentum 
$\vek r_f,\vek p_f$ are determined, for all particles, then these can be imaged back on  \emph{unique} classical trajectories to the initial $\vek r_i,\vek p_i$
of each particle \cite{footnote}.
In contrast to this classical behaviour, if less than complete measurements required to isolate a unique classical trajectory are made, then quantum interference can manifest itself in the measurement. This is analogous to ``which-way" experiments on double-slit interference. The specification of vectors 
$\vek r_f,\vek p_f$  is equivalent to a determination of which slit is traversed. Less information implies that a unique trajectory is not specified and interference can occur.

\begin{figure}[b]
\includegraphics[scale=.22]{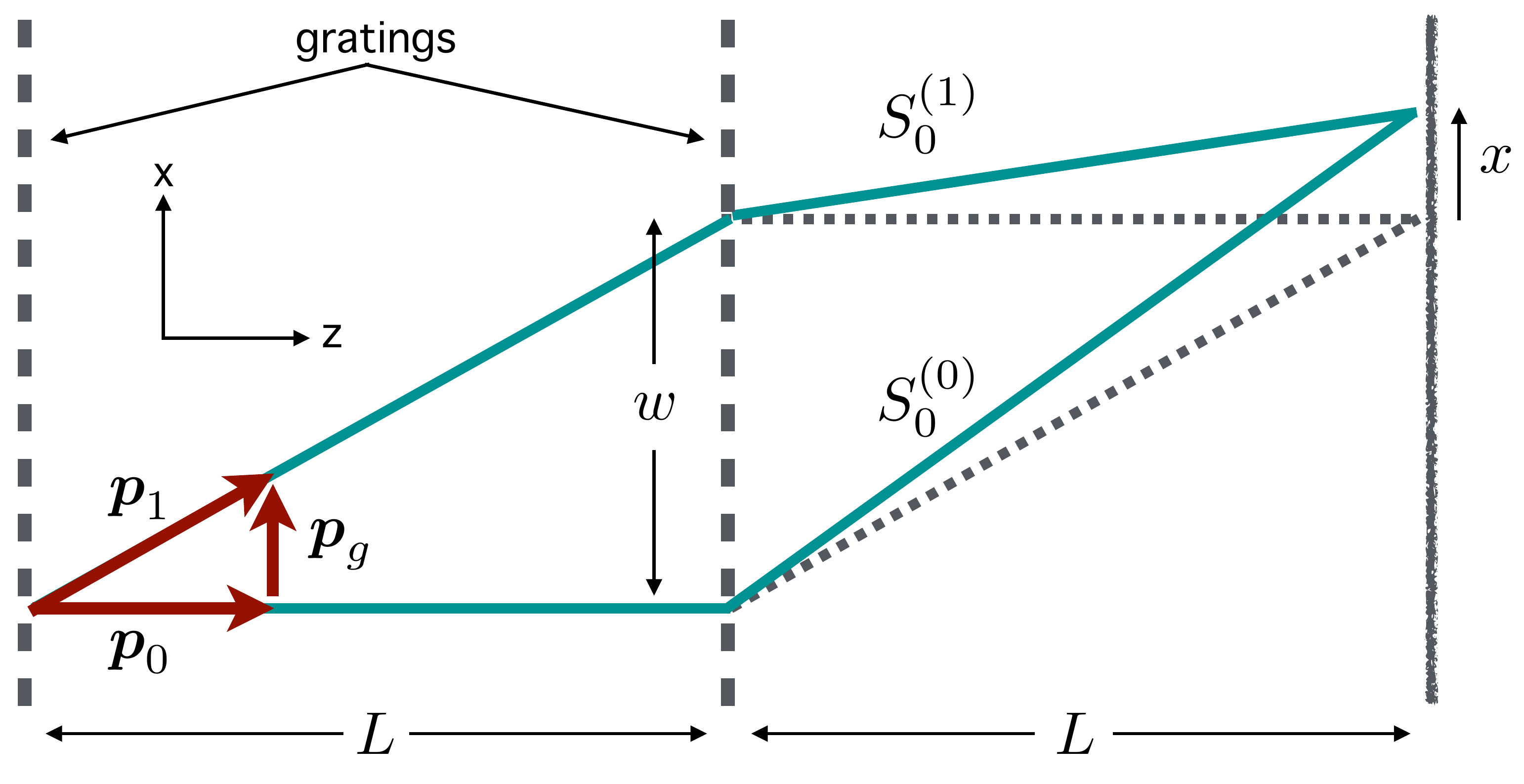}
\caption{\label{fig3} Mach-Zehnder atom interferometer \cite{Pritchard, Feagin2006} showing zeroth- and first-order diffraction characterising a pair of classical atom trajectories defined by momenta $\vek p_0$ and $\vek p_1= \vek p_0 + \vek p_g$ and the free-atom actions $S_0^{(0)}$ and $S_0^{(1)}$. }
\end{figure}

As example, consider an atom interferometer  \cite{Pritchard} constructed using a pair of nanofabricated gratings, as depicted in Fig.\ \ref{fig3}. The gratings consist of a line of $N$ equidistant narrow slits which are oriented along the $x$ axis and perpendicular to an incident beam of atoms along the $z$ axis. 
Just beyond the first grating, a single atom of mass $m$ and momentum $\vek p_0 = p_0 \hat{\vek z}$ is described with the wavefunction
\begin{equation}
\label{PsixGrating}
\Psi(\vek r, t_i) = (2\pi)^{-3/2} \, e^{i p_0 z/\hslash} F(x,y),
\end{equation} 
where $F(x,y)$ is the grating transmission (square-wave) function with grating period $d$ along $x$. The Fourier transform of this function is familiar \cite{BornWolf}, and the momentum wavefunction is easily obtained to give
\begin{equation}
\label{PsipGrating}
\tilde \Psi(\vek p) \propto \delta(p_0 - p_z) \, \frac{\sin N p_x d/(2\hslash)}{\sin p_x d/(2\hslash)},
\end{equation} 
where constant factors and a relatively broad single-slit diffraction function of $p_x, p_y$ have been dropped.
For $N$ large, the remaining multi-slit interference factor effectively vanishes unless $\vek p = \vek p_0 + \vek p_g n \, (n = 0, \pm1,...)$, where $\vek p_g \equiv (2\pi \hslash/d) \, \hat{\vek x}$ is a very small momentum transfer along the grating. In terms of the deBroglie wavelength $\lambda$ of the atom, $ p_g/p_0 = \lambda/d \ll 1$. The close analogy with optical interferometry is evident. 

One recombines the zeroth and first diffraction orders and thereby generates atom interference by inserting downstream the second grating parallel to the first. To achieve sufficient separation $w$ of the two corresponding atom trajectories at the second grating, the distance $L$ between gratings, which is also the distance from the second grating to the point where the two orders recombine, must be macroscopically large $L \gg w$.
Note $ p_g/p_0 = w/L$ and Fig.\ \ref{fig3} is not drawn to scale. In the experiments of Ref.\ \cite{Pritchard} $\lambda = 16 \,\mbox{pm}$, $d = 400 \, \mbox{nm}$, $L = 66 \, \mbox{cm}$, and $w \sim 30 \, \mu\mbox{m}$.

To establish interference fringes, consider the pair of nearby classical trajectories reaching a point $\vek r_f$ near the intersection of the two diffraction orders along a detection screen parallel to the gratings, as depicted in Fig.\ 3 \cite{Feagin2006}.
Along either trajectory, the atom's motion is described by the $3D$ free-particle classical action $S_0$ with the same time of flight,  $t = t_f - t_i = 2L/(p_0/m)$, along both trajectories.  The generalised IT \eref{Psixftf} gives the macroscopic wave function along each leg of the two trajectories. 
Thus the superposition describing the trajectory pair is given by 
\begin{equation}
\label{PsiMZ}
\begin{split}
\Psi(\vek r_f, t_f) \propto  \left (\frac{m}{it} \right)^{3/2} &\left[e^{i S_0^{(0)}/\hslash} \,  \tilde \Psi(\vek p_0)  \right. \\
	&+ \left. e^{i S_0^{(1)}/\hslash} \, \tilde \Psi(\vek p_0 + \vek p_g) \right],
\end{split}
\end{equation} 
where the classical density of free trajectories is given by $(m/t)^{3} = (p_0/(2L))^{3}$.
A short calculation shows that $S_0^{(0)} - S_0^{(1)} = p_g x = 2\pi \hslash \, x/d$, 
while for $N$ large $\tilde \Psi(\vek p_0) \approx \tilde \Psi(\vek p_0 + \vek p_g) \propto N$, so that 
the probability density describing the atom interference fringes as a function of $x$ is then given by   \cite{Feagin2006}
\begin{equation}
\label{fringes}
|\Psi(\vek r_f, t_f)|^2 \propto  \left(\frac{p_0}{2L} \right)^{3} 2N^2 \left[1 + \cos \left(\frac{2\pi }{d}x \right) \right]. 
\end{equation} 
In the experiments of Ref.\  \cite{Pritchard}, strong  fringes were observed in agreement with the IT predictions Eqs.\ (\ref{PsiMZ}) and (\ref{fringes}).

\subsection{The IT and decoherence}
It has become widely accepted that the reason we observe a classical world, although the motion of particles is governed by a quantum description, can be attributed to the phenomenon of decoherence \cite{decohere1,decohere}. This is the change in the wave function of a quantum system due to interaction with a quantum environment, variously taken to be the ambient surroundings, a measuring apparatus, or a combination of both. The necessary condition to achieve a classical status is considered to be the suppression of off-diagonal density matrix elements, leaving only diagonal elements, taken to indicate a transition from a quantum coherent superposition of amplitudes to a classical incoherent superposition of probabilities.

In the decoherence scenario, the deterministic propagation of the quantum system under the sole influence of its own Hamiltonian,
the ``von Neumann" term is  considered not to lead to classical behaviour.  Typically, the equation for the time and space propagation of the density matrix is split into three contributions \cite{decohere1,decohere}, namely, the von Neumann  term plus two phenomenological terms arising from interaction with an environment. The first is a dissipative term, whose influence on the quantum to classical transition is usually considered negligible. The important contribution is a stochastic, temperature-dependent interaction leading to decoherence. This is considered the driving term of the emergence of classical behaviour.  The loss of quantum coherence is signified by the off-diagonal elements of the density matrix in the basis appropriate to the measurement, e.g.\ the position basis, becoming zero. The only classical aspect is this loss of quantum coherence. However, the resulting particle dynamics are not proven to be Newtonian \cite{decohere1,decohere}.

The hermitian von Neumann term  is considered to describe deterministic purely quantum propagation unconnected with emergence of classical attributes except in so far as to point out that this part of the density matrix, or equivalently the Schr\"odinger  wave function, obeys Ehrenfest's theorem. However, this theorem shows merely how quantum averages, i.e.\ \emph{expectation values}, vary in time. For wave functions delocalised over macroscopic distances as considered here, the variation of expectation values is of little practical meaning. 

The IT demonstrates explicitly that the diagonal elements of the density matrix assume a classical form after  propagation over distance and times which are microscopic. This occurs even in a perfect vacuum. In fact
the IT shows that as soon as quantum particles leave a volume of microscopic dimensions in which their  accumulated action has become much greater than $\hslash$, their probability of detection is identical to the probability of the system being launched effectively from position $\vek r_i = 0$ with an initial classical momentum $ \vek p_i$ which ordains the system to arrive at a macroscopic  (detection) position $\vek r_f$ with momentum $ \vek p_f$ decided by the classical trajectory connecting these phase-space points. Detection of  individual particles at different macroscopic distances and times will lead an observer to infer a particle trajectory according to classical mechanics.

 The probability density of a position measurement is simply the diagonal element $\rho(\vek r_f,\vek r_f, t_f)$ of the density matrix in position representation. According to \eref{VVtheorem2} this is written
\begin{equation}
\rho(\vek r_f, \vek r_f, t_f) = |\Psi(\vek r_f, t_f)|^2 = \frac{d \vek p_i}{d \vek r_f} \, |\tilde\Psi(\vek p_i, t_i)|^2.
\end{equation}
and again we emphasise that the dynamics for varying time are purely classical; the quantum wavefunction merely providing an initial momentum probability distribution. One also notes that, since the position 
$\hat {\vek r}$ and momentum $\hat {\vek p}$ operators are both diagonal in the position representation,  any averages Tr$(\hat\rho \, \hat {\vek r})$, Tr$(\hat\rho \, \hat {\vek p})$ also only involve the diagonal elements of the operator $\hat\rho$. Despite the classical behaviour of contributions to the diagonal elements from individual trajectories, as shown in the previous section, measurements involving coherent contributions from more than one path will reveal full quantum interference. 
However, since interference still involves only the diagonal elements of $\rho$, as in \eref{fringes}, this is no different from the same phenomenon occurring with classical light or with coupled mechanical oscillators \cite{Alex}, for example. \\
Although, for the measurements considered here, the off-diagonal elements of the density matrix are not relevant, in view of their importance to decoherence theory we should examine their behaviour. These elements are defined
\begin{equation}
\rho(\vek r_f,\vek r_f',t_f) = \Psi^*(\vek r_f,t_f)\Psi(\vek r_f',t_f) 
\end{equation}
and from the form of the asymptotic IT wavefunction of \eref{Psixftf} and \eref{SemiKc} will contain oscillatory action phase factors varying in time and space.

To see the form of the IT density matrix elements in detail, again it is simplest to refer to the free motion of a single particle in one dimension from Sec.\ III.A. Here it is important to recognise that we consider the limit where both $z_f$ and $t = t_f - t_i$ becomes macroscopically large but their ratio is the constant classical velocity $v \equiv z_f/t$. Then, using \eref{PsiFreeIT},  the diagonal matrix element of $\rho$ can be written
\begin{equation}
\rho(z_f,z_f,t) = \frac{1}{\sqrt\pi \, V t}~e^{-v^2/V^2}
\end{equation}
where we define the constant ``velocity" $V \equiv \hslash/(m\sigma)$. Clearly the monotonic $1/t$ dependence conforms to the curves shown in Fig \ref{fig2}.

By contrast the off-diagonal elements are
\begin{equation}
\rho(z_f,z_f',t) = \frac{1}{\sqrt\pi \, V t}~e^{-(v^2 - v'^2)/(2V^2)} \, 
	e^{-i(v^2 - v'^2)t/(2V\sigma)}
\end{equation}
and clearly show oscillatory behaviour in time. Since all quantities involved have values of atomic size, the period $2\pi/\Omega$ with $\Omega \equiv (v^2 - v'^2)/(2V\sigma)$ takes values typically of the order of $10$ to $100$ a.u.\ of time $ \sim \! 10^{-17}\,\mbox{s}$, and hence one requires femtosecond resolution to observe such oscillations. For typical laboratory resolution of nanoseconds the off-diagonal elements will average to zero. Hence, whilst decoherence undoubtedly arises when external stochastic interactions are present, the IT, embodying the result of purely unitary Schr\"odinger propagation, is a vital accessory to the transition leading to the inference of classical behaviour of observed particles.

 If two classical trajectories contribute then the diagonal elements will also require appropriate resolution, in time or equivalently in space, to observe interference. 
In this case the off-diagonal elements of $\rho$ will be a sum of four oscillatory terms of differing  phase and thus will average to zero unless extremely high resolution is applied.
In the case of the atom-interferometry experiments \cite{Pritchard} considered in the previous section, the observed period of the fringes is $400 \, \mbox{nm}$, and it is necessary that the relative separation $L \sim 1 \, \mbox{m}$ of the gratings be stationary to $\sim \! 100 \, \mbox{nm}$ to observe any fringe contrast.

\section{Conclusions}
Using known results of semi-classical quantum mechanics we have generalised the IT to describe arbitrary motion of particles emanating from a microscopic reaction zone to a detector at macroscopic distance away. Although the motion is according to the Schr\"odinger equation, the space variables of the asymptotic wavefunction vary in time according to classical trajectories. Most importantly, the asymptotic behaviour is reached at times and distances which are on a microscopic scale. This result justifies the standard practice of experimentalists to use classical  mechanics to propagate measured position and momentum values back to the reaction zone. It also justifies the interpretation of many aspects of ionisation processes in strong laser fields, e.g. the generation of high harmonics, in terms of classical electron trajectories \cite{High harm}.  However, since the IT provides a wavefunction, at the same time it allows quantum interference patterns to be explained in terms of interfering contributions of classical trajectories, typified by the example of section IIIB.  In this way the generalised IT gives a concrete mathematical explanation of the apparent dichotomy of well-defined classical trajectories being associated with a quantum wavefunction. 

We have shown that the diagonal elements of the quantum density matrix, in the case where a measurement specifies both vector position and momentum, have a  \emph{purely classical form} describing a distribution of classical trajectories. If less than complete information is measured then more than one trajectory may contribute. If the time or spatial resolution is sufficient, then the diagonal elements can show interference structure. 

The off-diagonal elements of the quantum density matrix generally contain oscillatory terms which average to zero unless a high-resolution measurement is carried out. Such an elimination of off-diagonal matrix elements is considered the hallmark of decoherence due to  interactions \emph{external} to the quantum system and the signature of the transition from quantum to classical. However, the IT is a result of unitary Schr\"odinger propagation devoid of external influence. 
Of course, any interaction with the environment will provide the additional changes in the quantum system ascribed to the decoherence phenomenon.

The emergence of classical trajectories from quantum waves of course bears great similarity to the far-field emergence of ray optics from wave optics. Also in that case the inference of ray trajectories or observation of optical interference depends upon the resolution of detection. In particular, the IT relation of the asymptotic space wavefunction to the initial momentum wavefunction is mathematically similar to the Fraunhofer diffraction limit of Fresnel diffraction.
Interestingly, the origin of the classical characteristics of the IT are to be found in the wave nature of quantum physics itself. It is the cancelling out of differing oscillatory terms arising from action phase functions which leads mathematically to the stationary-phase approximation which isolates individual classical trajectories. In short, any quantum system propagating from a microscopic region to a macroscopic observation point will exhibit the classical characteristics described here; the quantum world autonomously becomes classical.

\section*{Acknowledgements}
JF acknowledges the ongoing support of the Department of Energy, Chemical Sciences, Geosciences and Biosciences Division of the Office of Basic Energy Sciences.


\begin{thebibliography}{}
\bibitem{footnote} In practice, rather than measure $\vek p_f$, one can also measure of course $\vek r_f$ and the time of flight for each detector hit to determine $\vek r_i,\vek p_i$. See Ref.\ \cite{BriggsFeagin_ITII}.
\end{thebibliography}
\end{document}